# EMERGENCY CENTRE ORGANIZATION AND AUTOMATED TRIAGE SYSTEM

**Dan Golding, Linda Wilson and Tshilidzi Marwala**

*School of Electrical & Information Engineering, University of the Witwatersrand, Private Bag 3, 2050, Johannesburg, South Africa*

**Abstract:** The excessive rate of patients arriving at accident and emergency centres is a major problem facing South African hospitals. Patients are prioritized for medical care through a triage process. Manual systems allow for inconsistency and error. This paper proposes a novel system to automate accident and emergency centre triage and uses this triage score along with an artificial intelligence estimate of patient-doctor time to optimize the queue order. A fuzzy inference system is employed to triage patients and a similar system estimates the time but adapts continuously through fuzzy Q-learning. The optimal queue order is found using a novel procedure based on genetic algorithms. These components are integrated in a simple graphical user interface. Live tests could not be performed but simulations reveal that the average waiting time can be reduced by 48 minutes and priority is given to urgent patients.

## 1. INTRODUCTION

South African public hospital Accident and Emergency Centre (AEC) queues are notoriously long. Recent years have seen these hospitals formalizing their medical triage systems, whereby patients are sorted before seeing the doctor to prioritize care to those most urgent. The Cape provinces are beginning to standardize their approaches under the Cape Triage System (CTS) [1]. However, no such standardization exists in the majority of the country. Furthermore, CTS does not make use of technology. The power of modern Computational Intelligence (CI) techniques has aided many industrial and service processes in becoming more automated and uniform [2]. This paper proposes a proof-of-concept system that employs a wide variety of such techniques encompassing machine learning, expert systems and optimization to automate the process of medical triage and digitally aid the management of an AEC in general.

A novel Genetic Algorithm (GA) based approach is applied to the scheduling problem. To find the optimal queue sequence, two factors are considered, namely patient urgency and individual queuing time. Patient urgency is considered based on Triage Scores (TS) as defined by CTS [1]. A Fuzzy Inference System (FIS) is used to automate the calculation of the TS. To estimate how long an individual will wait, it is necessary to have an idea of how long each member ahead of him in the queue will spend with the doctor. This is found using a FIS that is constantly being adapted via reinforcement learning. These technologies are integrated in a user-friendly Graphical User Interface (GUI).

The following section provides a brief background into medical triage, the current systems and how CI techniques have been applied. An overview of the new system is then presented as section 3, before sections 4 through 7 detail its implementation by considering the TS calculations via a FIS, the time predictions via Fuzzy Q-Learning (FQL), the GA based scheduler and the GUI respectively. Section 8 presents tests and simulations that ratify the system's success. Section 9 provides a critical discussion of all aspects of the proposal before the paper is concluded.

## 2. BACKGROUND

### 2.1. Medical triage

Public hospitals rarely have the capacity to help patients as they arrive. Triage is the practice of prioritizing patients based on their need for immediate attention and chance of recovery to ensure a maximum number of recuperations [3]. South Africa has no national triage standards. The Western Cape hospitals conform to CTS but their Gauteng counterparts have yet to implement such a framework. Few of the province's hospitals use any form of triage, which leads to dangerous queue lengths. The Johannesburg General Hospital (JGH)[1] has implemented a triage system based on CTS, explained in section 2.3. The system developed in this paper is based on the JGH but easily generalizes to any hospital.

### 2.2. Current solutions

CI medical triage applications are relatively new. A FIS battle-field triage system employs a similar inference technology to this paper [4]. The Dynasty Triage Advisor is an advanced system that uses Bayesian probability to match symptoms with diseases [5]. An automated triage and hospital check-in system, developed in Canada in 2007, sorts patients through hard coded rules [6]. However systems that use CI inferences to optimize hospital queues appear absent from publically available literature. Similar problems in factories and job-shops have attracted the attention of CI, particularly in the form of stochastic optimization such as GA [7]. Hospital queues do not compare in complexity to these problems and so the application of CI optimization seems to be the natural progression of research in automated triage.

---

[1] Now the Charlotte Maxeke Johannesburg Academic Hospital.

## 2.3. Triage at the Johannesburg General Hospital

Triage in the AEC of the JGH is performed by nurses. It requires rapid, complex calculations; a task which demands extensive training [8] that heavily burdens hospital resources. The following is a description of their triage process (CTS can be consulted for further detail [1]):

1. A nurse measures the patient's vital physiological parameters. The most essential of these are Systolic Blood Pressure (SBP), Heart Rate (HR), Temperature (T°) and Respiration Rate (RR).
2. Each of these vitals is scored using CTS. Table 1 shows the CTS scores for these vitals.
3. These scores are then summed and the total defines the triage colour. The nurse may then consider other ailments such as Per Vagina Bleeding (PVB) or localized pain and adjust the colour as necessary.
4. This information is recorded on paper by the nurse. The patient then enters the back of the queue unless they are truly urgent.

Table 1: Abridged CTS triage score table.

| TS  | 2     | 1      | 0        | 1       | 2       |
|-----|-------|--------|----------|---------|---------|
| SBP | 71-80 | 81-100 | 101-199  |         | >199    |
| HR  | <40   | 41-50  | 51-100   | 101-110 | 111-129 |
| T°  | <35   |        | 35-38.4  |         | >38.5   |
| RR  | <9    |        | 9-14     | 15-20   | 21-29   |

## 3. OVERVIEW OF SOLUTION

This section presents the system in its entirety to contextualize the CI technologies employed. Figure 1 provides a high-level flow chart of the system. The figure only represents the core functionalities; peripheral features, such as its search capabilities, are omitted.

The system serves to automate and optimize the triage and queuing processes and employs a GUI to manage this. Two forms, currently filled in by hand, are digitized. The first is the nurses' triage form and the second is the doctors' primary assessment form. The system's CI components are integrated into these two forms.

Once the nurse form is complete, a simple button facilitates the inference of the TS and a prediction of the time the patient will spend with the doctor. The patient's information is then submitted to a database and the queue.

A simple button on the doctors' form submits the doctor's comments, optimizes the queue, loads the information of the next patient to be seen and updates the time prediction model via reinforcement learning.

The following three sections detail the CI processes that form the core of this work.

## 4. FUZZY INFERENCE SYSTEM

FIS form a CI paradigm that performs regression like inferences in a manner similar to human reasoning [9]. Inputs and outputs are assigned partial memberships to broad classes. For example, a SBP input of 100 might fall in both the *normal* and the *low* classes. Semantic rules match inputs classes to output classes. This method is based around the concepts of fuzzy sets and logic and is described in detail in [10].

### 4.1. Motivation

The collection of prospective triage data from hospitals has proved to be logistically impossible. The only retrospective data available from the JGH was highly limited and so riddled with inconsistencies that a supervised learning approach was not viable. This left two options, an online learning approach or an expert system.

An online approach is preferable but troublesome to test. Thus online learning was reserved for predicting the time patients spend with the doctor as it provides sufficient proof of concept that such a system could be implemented for inferring TS at a later stage. This left expert systems. Communication with the JGH heads of triage and a review of the CTS revealed that a FIS is appropriate as it uses a rule base but can provide higher precision than CTS. Furthermore, inputs such as pain are already in a fuzzy form and so fit naturally into a FIS. In addition FIS have a low computational complexity and are simple to implement and interpret [9].

### 4.2. Implementation

The implementation of the FIS can be split into 3 design choices. First, the input space must be defined, then a set of Membership Functions (MF) is chosen for each input and output and finally a rule base is developed. After consultation with doctors responsible for triage at the JGH and in conjunction with the CTS the inputs are as follows:

- Vitals – SBP, HR, T°, RR
- Level of consciousness
- Pain – excluding in the limbs

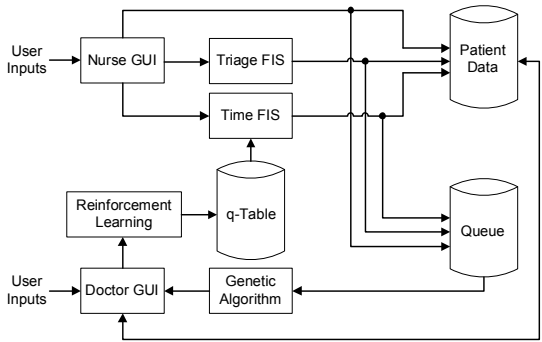

Figure 1: System overview.

MF for the vitals are based on the CTS ranges, similar to Table 1. The curve shapes are Gaussian with plateaus whose centres and spreads are adjusted in an ad hoc manner to optimize the output. Regional pains are given a score based on location and severity. The sum of these scores is used as an input to fuzzy MF developed in a similar manner. The output MF are based on CTS.

The rule base is designed around CTS in conjunction with the FIS MF. Additional rules for pain are based on expert opinions. Further rules from CTS are added after the FIS on an *if-then* basis. For example, *if* the output is green and the patient suffers from PVB, *then* the output is yellow.

Unlike CTS, the output is not left in a fuzzy form. Thus this system can differentiate severity between patients whom CTS classifies as *green*. This increased precision improves the ability of the scheduling algorithm.

## 5. FUZZY Q-LEARNING

Reinforcement learning is the unsupervised process whereby CI agents learn to act optimally in an uncertain environment [11]. The agent performs actions by trial and error and interprets environmental reaction as reward or punishment. Decision making parameters are then altered accordingly. FQL is a specialization of reinforcement learning where the decisions are made by a FIS and the parameters altered are the weights of the rule base.

*5.1. FQL algorithm*

FQL is a complex mathematical procedure and its details can be found in [11]. This section provides a short high level explanation of what FQL is conceptually.

FQL revolves around the q-table, a table of weights representing all possible rules connecting the input space to the output space. Its rows are the antecedents and columns the consequents. In fuzzy logic, an input vector can activate many antecedents [9]. FQL chooses the best consequent for a given set of antecedents. This is done by an Exploration-Exploitation Function (EEF). Exploration means choosing a consequent at random, exploitation means choosing the consequent with the highest value on the q-table. The EEF defines the probability of the consequent being chosen by either exploration or exploitation. An action is then performed using this consequent and the environment responds with a reward which is used to update the weight of the chosen consequent.

*5.2. Motivation*

No data are available on which to build an expert system or train a model by supervised learning. However it does seem plausible that the time spent with a doctor is somewhat influenced by measureable factors. The power of FQL is that if there are rules governing the behaviour of this system, they will be found.

The online learning capacity means that non-stationary processes, which this might well be, can be modelled. Furthermore, different systems can be trained to model different doctors.

*5.3. Implementation*

To ascertain what inputs should be considered, and if the model should be inference based or stochastic, would require live testing in a hospital environment. This was a logistical impossibility for this study and so inputs were arbitrarily selected to be the patient severity and the patient age. This selection is inconsequential as adding inputs is trivial. The EEF, tweaked by the tests in section 8.2, is described by (1)

$$\varepsilon = \begin{cases} 1 - 0.0038t & \text{when } t < 250 \\ 0.05 & \text{when } t \geq 250 \end{cases} \quad (1)$$

Where: $\varepsilon$ = probability of exploration
$t$ = epoch

Note that the probability of exploitation is thus *1 - ε*. Hence the policy is to begin with a high probability of exploration, allowing the system to find global optima, and linearly decrease this to a 5 % chance by epoch 250.

## 6. SCHEDULING ALGORITHM

Scheduling is the prioritizing, timing, and sequencing of work [12]. Said another way, scheduling is the process of finding an optimal sequence. Many algorithms are available, from deterministic to stochastic, single line to multi-line, simple priority based to complex orders for job-shops. Prudent use of scheduling can drastically reduce waiting times, save money and increase operational efficiency. It applies equally to manufacture as to service industries, the latter being our concern. Examples range from algorithms governing call centres to the ordering of instructions in digital processing [13]. Cleary AEC queues can draw from the benefits of scheduling.

*6.1. Constraints and considerations*

In a hospital queue, the work to be sequenced is the waiting patients. Each member has two factors that need to be considered: urgency and waiting time. Together these factors determine patient priority and thus can be used to define the optimal sequence. Both are essential. Irreversible damage can occur if urgent patients are made to wait too long. However, patients that are less urgent can not be expected to wait forever and thus the time spent in the queue must be considered.

Unlike job-shops where certain jobs have to be done before others are possible, hospital queues can be arranged in any permutation. This means that for a queue of 20 patients, a reasonable number for the JGH, the search space comprises approximately 2.5 x $10^{18}$ sequences. This makes the use of stochastic approaches necessary. GA was initially chosen for its ability to deal with sequencing natively [14]. However, the use of a novel algorithm that maps sequences to numbers means any stochastic optimization can be used. GA is chosen for its simplicity and widely available programming toolboxes.

GA is a stochastic optimisation technique based on biological evolution. The concept of *survival of the fittest* is used by modelling a population's fitness after the function to be optimized, which is hence known as the fitness function. The individual members of the population (each of which is a potential solution) undergo the biological processes of selection, recombination and mutation to form new generations which converge on global optima. Reference [15] provides a detailed explanation of GA.

*6.2. Novel Algorithm*

*Fitness function:* A novel fitness function is used that accounts for patient urgency as well as waiting times and is described by (2).

$$\sum_{i=1}^{n}\left((T_i+1)(t-t_{ai}+\sum_{k=1}^{i-1}t_{ek})\right) \quad (2)$$

Where:  $n$ = number of patients in queue
 $T_i$ = TS of patient $i$
 $t$ = current time
 $t_{ai}$ = time that patient $i$ arrived
 $t_{ek}$ = time that patient $k$ is expected to spend with the doctor

This equation assigns a value to a queue order by multiplying each patient's urgency with the total time they spend waiting. $T_i+1$ is the TS, i.e. patient urgency. The additional one ensures that the time information is not lost in the case where the TS is zero. The total time is broken up into the time the patient has already waited, $t-t_{ai}$, and the time the patient will still wait i.e. the sum of the time every person ahead in the queue is expected to spend with the doctor. Thus the goal of the GA is to find the sequence for which (2) yields the lowest result. One drawback of the GA is speed. The algorithm takes around 30 s

Table 2: The first 8 (of 24) permutations as seen by the mapping algorithm given $n=4$.

| Index | Sequence | | | |
|---|---|---|---|---|
| 1 | 1 | 2 | 3 | 4 |
| 2 | 1 | 2 | 4 | 3 |
| 3 | 1 | 3 | 2 | 4 |
| 4 | 1 | 3 | 4 | 2 |
| 5 | 1 | 4 | 2 | 3 |
| 6 | 1 | 4 | 3 | 2 |
| 7 | 2 | 1 | 3 | 4 |
| 8 | 2 | 1 | 4 | 3 |

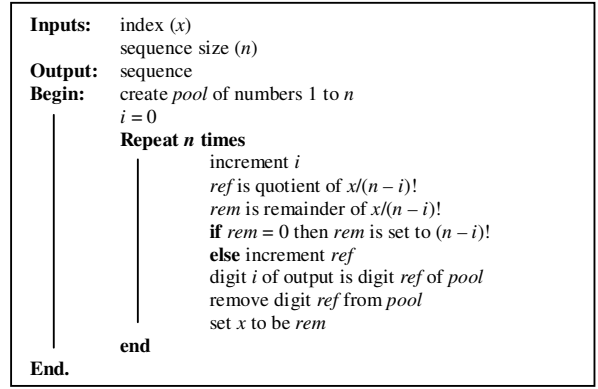

```
Inputs:    index (x)
           sequence size (n)
Output:    sequence
Begin:     create pool of numbers 1 to n
           i = 0
           Repeat n times
                   increment i
                   ref is quotient of x/(n – i)!
                   rem is remainder of x/(n – i)!
                   if rem = 0 then rem is set to (n – i)!
                   else increment ref
                   digit i of output is digit ref of pool
                   remove digit ref from pool
                   set x to be rem
           end
End.
```

Figure 2: Pseudo-code of the mapping algorithm.

for an exceptionally long queue of 100 members. Whilst this is slow, it is perfectly reasonable when considering the application, after all it will take a patient longer than 30 s to leave the doctors room.

*Mapping function:* GA generates a population of numbers, not sequences. Thus a function is required to map these numbers to all possible permutations of a given queue length. The function arranges permutations so as to minimize change between permutations of consecutive indices. This arrangement leads to a smoother function which is easier to optimize [16]. Table 2 shows a sample of permutations and there corresponding indices.

The GA generates indices but the fitness function requires sequences. An algorithm that finds a sequence given an index and the number of elements has been developed and is described by Figure 2. It is based on the fact that numbers are grouped in batches of $p!$ where $p$ is $n$ minus the element number. For example, notice the group of six ones in the first column of the sequences in Table 2. This is element one where $n=4$ thus $p!=6$. Hence the quotient of dividing the index by $p!$ is one less than the number for that element. The following elements are found through the same process, but instead of the index the remainder is divided. The only exception occurs when there is no remainder. In this case it is not necessary to add 1 and $p!$ is considered the remainder instead of zero.

## 7. INTERFACE AND IMPLEMENTATION

*7.1. Interface*

Whilst not core to the solution, the GUI demonstrates how these complex and powerful CI tools can be integrated into a simple interface. The GUI has two main forms, shown in Figure 3. The forms are designed to be as user-friendly as possible. It is essential that the use of an electronic form does not hinder the nurse in any way. Thus the form is designed to be quick and simple to use, and eliminates the need for free typing wherever possible.

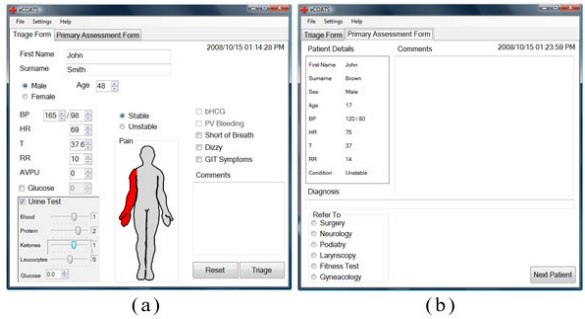

Figure 3: Screen shots of the GUI. The nurse form is shown in (a) and (b) shows the doctor form.

The pain interface simply requires the nurse to click on the anatomical region where the patient feels pain. This turns yellow to indicate mild pain, and another click turns it red for severe pain.

The doctor's form automatically provides information such as patient name and vitals. The doctors' notes are now typed eliminating their notorious problem of illegible hand-writing. The *Next Patient* button loads this information and triggers the necessary CI, which tells the doctor who is next from the re-optimized queue.

*7.2. Development*

Three languages have been used in the system's development. The GUI is programmed in *C#*. This language is designed for developers to be able to produce applications in minimal time. Being object oriented, it allows for highly modular programming facilitating future expansion. The information is stored using *MsSQL*. *SQL* is a tried and tested databasing language which can be accessed natively in *C#*. It is used for reliability and ease of development. Finally, the technical CI components, the core of the system, are developed in *Matlab* as it gracefully handles complex numerical calculations. Furthermore, *Matlab* has toolboxes that cater for FIS and GA and has a powerful visualization functionality which is a tremendous aid when testing via abstract simulations.

## 8. TESTS AND SIMULATIONS

Time and logistics do not allow for a thorough, live testing of the system. For this reason a series of simulations are used to test each CI aspect individually.

*8.1. Triage score inference*

Simulations are based on data collected from the JGH. Comparing results of the FIS with those from the JGH is futile due to the inconsistencies of the nurses, so they are compared with CTS instead. Pain is considered separately as CTS isn't specific about it. Success is measured against common metrics for under-triage (less than 5 % of cases) and over-triage (less than 50 % of cases) [17]. Table 3 shows the results of these tests.

Table 3: Percentage of under and over triage.

|  | Under Triage | Correct Triage | Over Triage |
|---|---|---|---|
| **No Pain** | 2 | 96 | 2 |
| **Low Pain** | 1 | 91 | 8 |
| **Medium Pain** | 0 | 60 | 40 |
| **High Pain** | 0 | 37 | 63 |

*8.2. Prediction of time to be spent with doctor*

Neither literature nor data is available on these predictions. This test is purely a proof-of-concept. However, it is possible that no relationship determines how long doctors spend with patients, or that the relationship is stochastic in which case statistical modelling might be preferable. It is not possible to know this without running tests in a live hospital environment.

It is assumed that there is at least some relationship governing this time and that it can be inferred from symptomatic and demographic information. To test the online learning capabilities, random FIS models are made and data are produced using these models. The system is then trained and the similarity between the training system and the online learning system is investigated.

In all cases, the FQL model assigns its highest values on the q-table to the rules of the FIS it is learning from. Tests reveal that an EEF which reduces exploration to a probability of 5 % after 200 epochs produces the best results. Figure 4 shows a sliding average (over 100 epochs) of the time differences between the simulation and the FQL model as it learns. Once the model has settled, the average absolute difference is less than 4 min.

*8.3. Scheduling*

The data collected form the JGH provides the times that patients arrive and the time that they see the doctor. Unfortunately the time spent with the doctor is not recorded. It is assumed that the consecutive times between when patients see the doctor indicate the time spent with the doctor. To simulate the output of the online learning process, noise is added to these times as characterized by the tests in section 8.2. Figure 5 shows the waiting times of a queue of 17 patients with and without scheduling. The average waiting time from the JGH data is 169 min. Scheduling reduces this to 121 min.

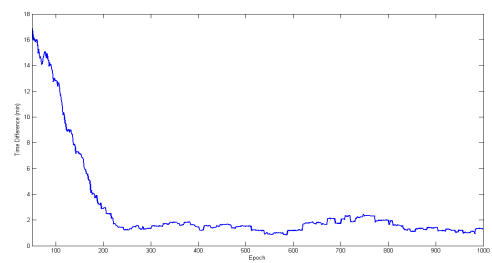

Figure 4: Time difference of a FIS training a FQL model.

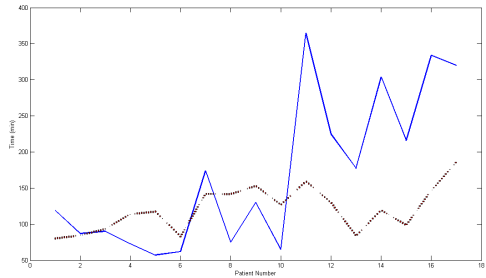

Figure 5: Waiting time without scheduling (solid blue line) and with scheduling (dashed black line).

## 9. CRITICAL ANALYSIS

*9.1. Evaluation*

The proposed solution has achieved the goal of being a proof-of-concept automated triage system. The CI technologies at the core of the system have been tested and show promising results. Simulations reveal that queue optimization can reduce the average patient waiting time by 48 minutes. The FQL system has proved to be able to learn from a rule based environment to predict the time a patient will spend with the doctor on average to within 4 minutes of the true time. The triage system has reduced under-triage to less than 2 %. Over-triage is more difficult to interpret as increased pain should increase the TS. Only in cases of severe pain does the FIS triage higher than CTS for more than 50 % of cases; however CTS is likely to under-triage in these cases which is why it recommends nurses prioritize patient with pain higher than their TS. The system serves as sufficient proof that such a product is worth testing in a live hospital environment.

*9.2. Future Work*

There is the potential to implement numerous useful features from automatic calibration of nursing equipment to enhancing the automated triage system with natural language processing. The most pressing work however is the running of live tests. The system needs minor alterations, mainly to do with networking, to be ready for testing in a hospital environment. Only with such tests can the CI components truly be evaluated and improved. Further work into the reinforcement learning aspects is also recommended but can not be done without live tests. Tests will show what inputs are appropriate for the time predictor. FQL should also be implemented in the triage prediction stage. Both FQL models should allow for the MF shapes to be trained in addition to the rule base.

## 10. CONCLUSION

CI tools can have a powerful impact on the running of AEC. Inference models remove nurse bias and human error from the triage process and scheduling reduces patient waiting time and increases efficiency of the hospital environment. The developed system shows promising results through simulations. Analysis of these results shows that this is a product well worth testing in a hospital environment to truly ascertain what impact such a system can have on the medical industry.

## ACKNOWLEDGEMENT

The author acknowledges Prof. Boffard and Dr. Motarra of the JGH for their help and advice regarding current triage systems in South Africa and accommodating the collection of triage data.